\documentclass[secnumarabic,oncolumn,graphics,floatfix,nofootinbib,
tightenlines,nobibnotes,aps,prl]{revtex4-1}

\usepackage{bm}
\usepackage{subfigure}\usepackage{color,amsmath,graphicx,epsfig,amssymb}

\newcommand{\bomega}{\boldsymbol \omega}
\newcommand{\bnabla}{\boldsymbol \nabla}

\newcommand{\gv}[1]{\ensuremath{\mbox{\boldmath$ #1 $}}} 
\newcommand{\grad}[1]{\gv{\nabla} #1} 
\newcommand{\br}{{\bf r}}

\newcommand{\bs}{{\bf s}}
\newcommand{\bv}{{\bf v}}

\begin{document}

\title{Helicity Conservation and Twisted Seifert Surfaces for Superfluid Vortices}

\author{
Hayder Salman$^{1}$}

\affiliation{$^{1}$School of Mathematics, University of East Anglia, Norwich Research Park, Norwich NR4 7TJ, United Kingdom\\
}



\email{H.Salman@uea.ac.uk}

\begin{abstract}
Starting from the continuum definition of helicity, we derive from first principles its different contributions for superfluid vortices. Our analysis shows that an internal twist contribution emerges naturally from the mathematical derivation. This reveals that the spanwise vector that is used to characterise the twist contribution must point in the direction of a
surface of constant velocity potential. 
An immediate consequence of the Seifert framing is that the continuum definition of helicity for a superfluid is trivially zero at all times. 
It follows that the Gauss-linking number is a more appropriate definition of helicity for superfluids.
Despite this, we explain how a quasiclassical limit can arise in a superfluid in which the continuum definition for helicity can be used. This provides a clear connection between a microscopic and a macroscopic description of a superfluid as provided by the Hall-Vinen-Bekarevich-Khalatnikov equations. This leads to consistency with the definition of helicity used for classical vortices.
\end{abstract}
\maketitle


\section{Introduction}

Many physical systems in nature support topological vortex excitations. These vortices correspond to phase singularities or defects that form in the order parameter that characterise the state of a given system \cite{Taylor2016}. Phase defects correspond to quantized vortices in Bose-Einstein condensates, superfluid Helium, optics, and superconductors. 
In contrast to vortices in a classical viscous fluid, these phase singularities correspond to filamentary vortex structures where rotation is concentrated entirely along a curve in 3D space. 

These filamentary defects can be arranged in complex and topologically non-trivial structures such as knots and links. This observation was conceived by Lord Kelvin \cite{Thomson1867} in his now discarded proposal of the vortex atom theory. Nevertheless, recent experimental advances that have allowed these structures to be realized in very different physical systems have revived interest in knotted and other topologically non-trivial vortices. While the work of \cite{Kleckner2013a,Kleckner2013b} produced classical knotted vortices in water, knotted phase defects have also been demonstrated in optical systems \cite{Berry2001,Dennis2010,Kedia2013}, and more recent works have focussed on atomic Bose-Einstein condensates \cite{Kawaguchi2008,Hall2016,Maucher2016a}.

The ability to create such knotted vortices under controlled conditions in the laboratory makes it possible to study what role topology plays in the long-time relaxation of a system towards equilibrium. In most systems, the topology is not frozen into the initial field because vortex reconnections allow the topology of vortices to change\cite{Maggioni2010,Proment2012}. Consequently, characterising how knotted vortices untangle allows us to understand different routes of relaxation of a given system \cite{Kobayashi2016,Kleckner2016,Liu2016}. In some systems, reconnections of vortex filaments are not permitted \cite{Maucher2016b}. In such cases, the lowest accessible energy state is determined by the type of knotted structures initially present.

It is clear that quantifying the degree of linkage of vortex lines is important to all of these problems.
A quantity that has been introduced in classical fluids to quantify this is the kinetic helicity \cite{Moffatt1969,Moffatt2014}. 
Helicity has an important significance because, for an ideal classical fluid in 3D (one that is perfectly destitute of viscosity), it is the only other known quadratic invariant besides energy. 
Given its importance, there have been attempts to quantify helicity for the filamentary vortex structures arising in a superfluid \cite{Scheeler2014,Leoni2016}.
However, adapting results of helicity conservation to such vortex filaments has been plagued with difficulties and apparent paradoxes. This has led several authors to conclude that helicity may not be conserved in such systems.

It is generally accepted that helicity is composed of several contributions among which internal (or intrinsic) twist makes up an important component and without it helicity conservation can not be ensured \cite{Moffatt1992}. However, it is not obvious how a structureless vortex filament can acquire an internal twist contribution. For this reason, several works have simply ignored such a contribution altogether with a consequent loss in helicity conservation \cite{Scheeler2014}. Other works have proposed defining the twist in terms of a surface of constant phase \cite{Zuccher2015,Hanninen2016}. This latter approach turns out to alleviate many of the inconsistencies that would otherwise arise by ignoring the twist contribution altogether. However, this choice is arbitrary and has not been justified from first principles. At present, there is no consensus on what is a correct definition of helicity for superfluid vortices. A related and equally important question is: how does a classical limit of helicity emerge from superfluid vortices? Indeed, in the classical limit, helicity should be conserved in the absence of vortex reconnections. These considerations reinforce the case for a correct definition of helicity for vortex filaments.

In this work, starting from the continuum definition, we will derive the different contributions to helicity of a superfluid vortex filament from first principles. Our derivation reveals how an internal twist contribution can arise. The results obtained clarify how a classical limit emerges that is fully consistent with known attributes of helicity in classical fluids.

\section{Helicity of Superfluid Vortices}

Superfluid vortices can be described as vortex filaments where all the vorticity is concentrated along a curve and where the density of the superfluid vanishes. If the system consists of $N$ closed vortex filaments (which will be assumed in the remainder of this work), then given a superfluid velocity field ${\bf v} ({\bf r},t)$, we can express the vorticity of the fluid ${\bf \bomega} = \nabla \times {\bf v}$ at the field point ${\bf r}$ as
\begin{equation}
{\bomega}({\bf r},t) =
\sum_{i=1}^N \Gamma_i \oint d\sigma_i \frac{\partial {\bf s}_i}{\partial \sigma_i} \delta^{(3)} \left( {\bf r} - {\bf s}_i(\sigma_i,t) \right). \label{eqn_vort}
\end{equation}
Here, $\sigma_i$ represents a parameterisation of filament $i$, $\delta^{(3)}$ is the Dirac delta function in three dimensions, ${\bf s}_i(\sigma_i,t)$ is the instantaneous configuration of vortex filament $i$ at time $t$, and $\Gamma_i$ corresponds to the circulation of the vortex. 

If $\sigma_i$ represents the natural paramerisation in terms of arclength, then ${\partial {\bf s}_i}/{\partial \sigma_i} = {\bf t}_i(\sigma_i,t)$ where ${\bf t}_i$ is the local unit tangent vector to the filament. The vorticity vector field is then always aligned along the filament and points in the direction that $\sigma_i$ is increasing.
To derive the different contributions to helicity, we begin with its definition \cite{Moffatt1969}
\begin{equation}
\mathcal{H} = \int {\bomega} \cdot {\bf v} d^3{\bf r},\label{eqn_Helicity}.
\end{equation}
where the integral is over the entire fluid domain. Adopting a Helmholtz decomposition, we can divide the velocity field into a divergent-free (${\bf v}_{_I}$) and a curl-free (${\bf v}_{_C}$) part
\begin{align}
{\bf v} = {\bf v}_{_I} + {\bf v}_{_C}, \;\;\;\;\; \text{where} \;\;\;\;\; \grad \cdot {\bf v}_{_I} = 0, \;\;\;\;  \grad \times {\bf v}_{_C} = {\bf 0}. \label{eqn_Helmholtz}
\end{align}
For a vortex filament, the incompressible velocity can be recovered from the Biot-Savart law given by
\begin{align}
{\bf v}_{_I}({\bf r},t) = \sum_{i=1}^N \frac{\Gamma_i}{4\pi} \oint_{\mathcal{C}_i} \frac{({\bf s}_i-{\bf r}) \times d{\bf s}_i}{|{\bf r}-{\bf s}_i|^3}, \label{eqn_BS}
\end{align} 
where $\mathcal{C}_i$ denotes the set of points lying along the $i$'th filament. We note from Eq.~\eqref{eqn_Helmholtz} that we can write ${\bf v}_{_C}= \grad \vartheta$ where $\vartheta$ is a smooth velocity potential. It follows that ${\bf v}_{_C}$ does not contribute to the helicity.

The integral in Eq.~\eqref{eqn_BS} diverges as ${\bf r} \rightarrow {\bf s}_i$ (i.e.\ if one evaluates the velocity on the vortex line given by ${\bf v} = {\bf v}({\bf s}_i,t)$). However, according to Eqs.\ \eqref{eqn_vort} and \eqref{eqn_Helicity}, only the tangential component of the velocity contributes to the helicity. It turns out that this tangential component remains well behaved because the divergent component of the velocity in the Biot-Savart integral is normal to the vorticity vector of the filament.

To evaluate the helicity, we can use Eqs.\ \eqref{eqn_vort} and \eqref{eqn_BS} for a vortex filament to obtain
\begin{align}
{\mathcal H} &= \int {\bomega} \cdot {\bf v} d^3{\bf r} = \int \sum_{i=1}^N \Gamma_i \oint_{\mathcal{C}_i} d\sigma_i \frac{\partial {\bf s}_i}{\partial \sigma_i} \delta^{(3)} [{\bf r}-{\bf s}_i(t)] 
\cdot \left[ \sum_{j=1}^N  \frac{\Gamma_j}{4\pi} \oint_{{\mathcal C}_j}  \frac{(\tilde{\bf s}_j-{\bf r}) \times d\tilde{\bf s}_j}{|{\bf r}-\tilde{\bf s}_j|^3} \right] d^3{\bf r} \, . \label{eqn_HelicityBS}
\end{align}
This would suggest that the helicity consists of only two contributions, the writhe (${\mathcal Wr}$), and the linking number ($Lk$), which are given by \cite{Pohl1968,White1969,Fuller1971}
\begin{align}
\mathrm{Lk} &= \sum_{\substack{i,j=1 \\ i \ne j}}^{N,N} \frac{\Gamma_i \Gamma_j}{4\pi}\oint_{\mathcal{C}_i} \oint_{\mathcal{C}_j} \frac{({\bf s}_i-\tilde{\bf s}_j) \cdot d{\bf s}_i \times d\tilde{\bf s}_j}{\left| {\bf s}_i-\tilde{\bf s}_j \right|^3} , \label{eqn_LinkNo} 
\end{align}
and
\begin{align}
{\mathcal Wr} &= \sum_{i=1}^N  \frac{\Gamma_i^2}{4\pi} \oint_{\mathcal{C}_i} \oint_{\mathcal{C}_i} \frac{({\bf s}_i-\tilde{\bf s}_i) \cdot d{\bf s}_i \times d\tilde{\bf s}_i}{\left| {\bf s}_i-\tilde{\bf s}_i \right|^3}. \label{eqn_Writhe}
\end{align}
We note that for the writhe, the integrand remains finite as ${\bf s}_i \rightarrow \tilde{\bf s}_i$ and so the apparent singularity appearing in the denominator of Eq.\ \eqref{eqn_Writhe} does not result in any divergences. 

This result stands in contrast to what is obtained if one considers a vortex tube (a region of fluid that is rotational within a finite core but irrotational outside). In particular, if we denote the centre-line of the vortex tube $i$ with the position vector ${\bf s}_i$, then, as shown in \cite{Moffatt1992}, the helicity is given by
\begin{align}
{\mathcal H} = (\mathrm{Lk} + {\mathcal Wr} + \mathrm{Tw}). \label{eqn_Helicitycont}
\end{align}
The linking number and writhe contributions to the helicity for the tube can still be expressed as in Eqs.\ \eqref{eqn_LinkNo}-\eqref{eqn_Writhe}. However, in this case, the helicity has a third contribution that corresponds to the twisting, Tw, of the vorticity within the vortex tube. 
Moffatt and Ricca \cite{Moffatt1992} showed that the twist arises from the meridional component of vorticity.
This internal twisting can be characterised by defining a ribbon where the boundaries are delineated by the centreline ${\mathcal C}_i$ and a nearby vortex line.
It is clear that such a definition prescribes the ribbon upto an arbitrary overall phase since different neighbouring filaments could be chosen.  To define the spanwise vector of the ribbon, we introduce the Serret-Frenet relations given by
\begin{align}
\frac{d {\bf t}}{d \sigma} = \kappa {\bf n}, \;\;\;\;
\frac{d {\bf n}}{d \sigma} = -\kappa {\bf t} + \tau {\bf b}, \;\;\;\;
\frac{d {\bf b}}{d \sigma} = -\tau {\bf n},
\end{align}
where ${\bf t}$, ${\bf n}$, and ${\bf b}$, $\kappa$ and $\tau$  are the unit tangent, unit principal normal, unit binormal vectors, the local curvature, and the local torsion, respectively.
Denoting the spanwise vector of the ribbon by ${\bf N}$, we have ${\bf N} = {\bf n} \cos \Theta + {\bf b} \sin \Theta$, where $\Theta=\Theta(\sigma,t)$ specifies the direction of the ribbon. We note that while the direction is defined upto an arbitrary phase, the twisting of the ribbon is encoded in the function $\Theta$. Following \cite{Moffatt1992,Banchoff1975}, it can then be shown that 
\begin{align}
Tw &= \sum_{i=1}^N \frac{\Gamma_i^2}{2\pi} \oint_{\mathcal C_i} ({\bf N}_i \times {\bf N}_i') \cdot {\bf t}_i d\sigma_i \nonumber \\
&= \sum_{i=1}^N \frac{\Gamma_i^2}{2\pi} \oint_{\mathcal C_i} \left( \tau_i + \frac{d \Theta_i}{d\sigma_i} \right) d\sigma_i = \sum_{i=1}^N  \Gamma_i^2 \left( \mathcal{T}_i 
+ \frac{1}{2\pi} \left[ \Theta_i \right]_{\mathcal C_i} \right),  \label{eqn_Twist}
\end{align} 
where a prime denotes partial differentiation with respect to arclength. According to \cite{Moffatt1992}, the quantity $\frac{1}{2\pi} \left[ \Theta \right]_{\mathcal C} \equiv \mathcal{N}$ provides a measure of the internal twist. 
The above considerations suggest that twist can be defined if the vortex has internal structure that results in twisted vortex filaments within the vortex tube with a finite core size. 

For a superfluid vortex filament, such an unambiguous choice for the spanwise vector appears to be lacking since, according to Eq.\ \eqref{eqn_vort}, a superfluid vortex has no internal structure. Indeed, while one can identify a vortex core with a superfluid vortex that is associated with the depletion of the superfluid density, the vorticity remains concentrated along a vortex filament. Therefore, the mechanism for generating twist in a classical vortex tube does not appear to apply to a quantized superfluid vortex. 
The apparent absence of the twist contribution has profound implications since as noticed in \cite{Scheeler2014} and further clarified in \cite{Hanninen2016}, helicity can not be conserved if a twist contribution is absent, even in the absence of vortex reconnections. 

However, contrary to the conclusions arrived at from these considerations, we will argue that an additional contribution to helicity can indeed be identified for a vortex filament that corresponds to twist. To accomplish this, we will work with the velocity potential formulation of a vortex filament. We note that the velocity potential is closely related to the phase of the wavefunction that describes the condensate order parameter.  For a superfluid, described by a complex scalar order parameter $\psi({\bf r},t)$, we can define the velocity and density of the superfluid as 
\begin{align}
\rho = m |\psi|^2, \;\;\;\;\; \text{and} \;\;\;\;\; {\bf v} = \frac{\hbar(\psi \grad \psi^* - \psi^* \grad \psi)}{i2m|\psi|^2} \equiv \frac{\hbar}{m} \grad \varphi,
\end{align}
where $\varphi$ is the phase of the wavefunction. 
The velocity potential we are interested in, which we denote by $\phi$, is given by the divergent-free projection of the velocity field, which corresponds to ${\bf v}_I = \grad \phi$. This potential will retain the phase singularities present in the phase of the wavefunction $\psi$. Hence, analysing the properties of this velocity potential will allow us to obtain a direct understanding of the properties of the phase of the wavefunction for a superfluid vortex. 

To motivate our approach, we recall that the velocity potential around a closed vortex filament has the property that
\begin{align}
\bomega = \grad \times {\bf v}({\bf r},t) = \grad \times \grad \phi, \label{eqn_irrvel}
\end{align}
where $\bomega$ is given in Eq.\ \eqref{eqn_vort}. These equations imply that away from the (possibly knotted) vortex filament where the flow is irrotational, a non-singlevalued velocity potential exists such that near the filament ${\mathcal C}_i$, $\phi \sim (2\pi)^{-1} \Gamma \theta$ and $\theta$ is the azimuthal angle measured in a plane normal to the local tangent vector of the vortex.
We can identify a distinguished surface $S(t)$ to be that where the (non single-valued) velocity potential jumps. This non single-valuedness arises from having a non-zero circulation associated with the vortex such that Eq.\ \eqref{eqn_irrvel} is satisfied. From the definition of the circulation given by $\Gamma = \oint_{\partial \Omega} \nabla \phi \cdot d{\bf l}$ that is evaluated along the circuit ${\partial \Omega}$ that is threaded by a vortex filament, it is clear that across the surface $S(t)$, we have the jump condition
\begin{align}
\Gamma = \left. \phi({\bf r},t)\right|_{S^+} - \left. \phi({\bf r},t)\right|_{S^-}. \label{eqn_jumpcondphi}
\end{align}

\begin{figure}[t]
\centering
\hspace{-1.cm}
 \begin{minipage}[b]{0.49\textwidth}
    \centering
        \includegraphics[width=3.in]{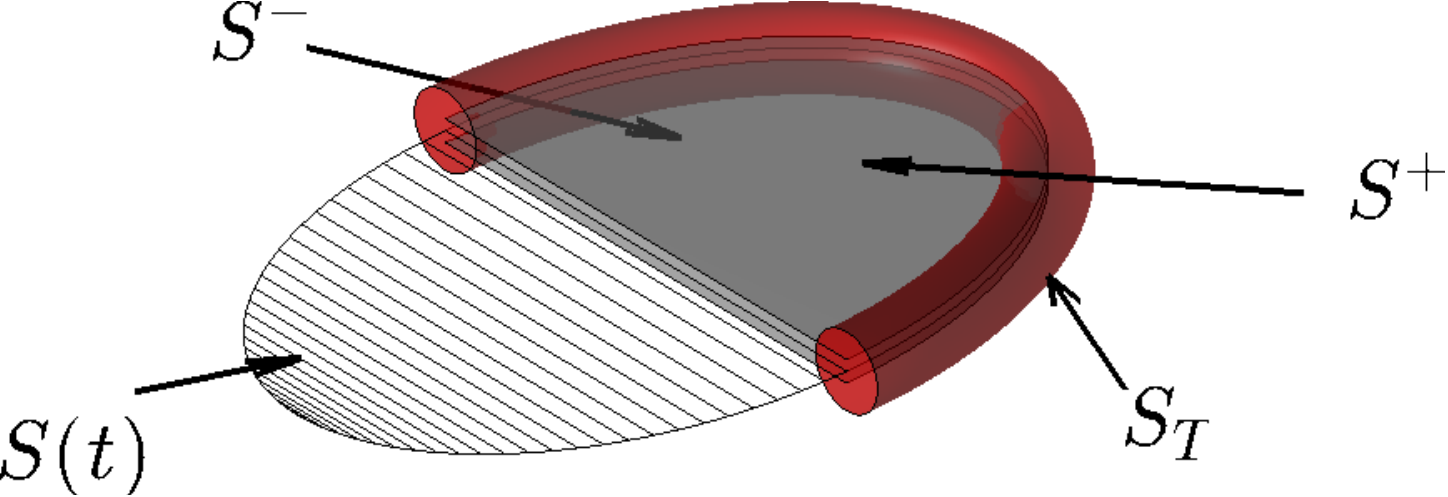}           
  \end{minipage}
 \vspace{0.2cm}
  \caption{Illustrattion of the distinguished Seifert surface(hatched region)  spanning the vortex filament, and the nearby surfaces of constant phase $S^+$ and $S^-$ across which there is a jump in the velocity potential $\phi$. The tube depicts the surface $S_T$ referred to in the text. \label{Fig_illustration}}
\vspace{-0.4cm}
\end{figure}

To evaluate the velocity potential of a closed vortex filament, we solve Laplace's equation 
\begin{align}
\nabla^2 \phi = 0, \label{eqn_Laplace}
\end{align}
that holds for an incompressible irrotational fluid subject to the boundary condition given by Eq.\ \eqref{eqn_jumpcondphi} and 
the far-field boundary condition $\phi \rightarrow \text{constant}$ as $|{\bf r}| \rightarrow \infty$.
Following the procedure in \cite{Lund1989,MilneThomson1968}, we start with Green's second identity
\begin{align}
& \int_{\Omega} \left[ (\phi (\tilde{\bf r}) \nabla_{\tilde{\bf r}}^2 G(\tilde{\bf r},{\bf r}) - G(\tilde{\bf r},{\bf r}) \nabla_{\tilde{\bf r}}^2 \phi (\tilde{\bf r}) )\right] d^3 \tilde{\bf r} \nonumber \\
=& \int_{\Omega} \left[ \grad_{\tilde{\bf r}} \cdot \left( \phi(\tilde{\bf r}) \grad_{\tilde{\bf r}} G(\tilde{\bf r},{\bf r}) \right) - \grad_{\tilde{\bf r}} \cdot \left( G(\tilde{\bf r},{\bf r}) \grad_{\tilde{\bf r}} \phi (\tilde{\bf r}) \right) \right] d^3 \tilde{\bf r}.
\label{eqn_Greens2id}
\end{align}
In general, this identity applies to any two sufficiently smooth functions $\phi$ and $G$. However, for our purposes, we will require $G$ to be the Green's function of Laplace's equation, which in 3D satisfies
\begin{align}
\nabla^2 G = 4\pi \delta^{(3)} \left( {\bf r}-\tilde{\bf r} \right), \;\;\;\;\; G({\bf r},\tilde{\bf r}) = G({\bf r}-\tilde{\bf r}) = \frac{1}{|{\bf r}-\tilde{\bf r}|}.
\end{align}
Since Green's theorem can only be applied to a simply connected domain where the functions $\phi$ and $G$ are sufficiently smooth, the region of integration, $\Omega$, appearing in Eq.\ \eqref{eqn_Greens2id} must exclude the surface $S(t)$ as well as the vortex filament. Such a region can be constructed by excluding a thin tube around the filament defined by the surface $S_T(t)$, and a thin slab around $S(t)$ that corresponds to the two surfaces $S^+$ and $S^-$ as shown in Fig.\ \ref{Fig_illustration}. The region denoted by $\Omega$ then consists of all points falling outside the excluded region and bounded by another outer surface, $S^o$, that can be taken to be a large sphere of radius $R$. To ensure the volume remains simply connected, the surface enclosing the excluded volume can be extended to the outer surface $S^o$ by cutting a thin volume away from the region bounded by the inner surface given by $\{S^i: S^i = S_T \cup S^+ \cup S^- \}$, and the outer surface, $S^o$. However, the contribution arising from these surface integrals vanish. We can, therefore, safely take $\Omega$ to be the entire volume of fluid lying outside the excluded region enclosing the filament and the surface $S(t)$ and bounded by the outer surface. It follows from Eq.\ \eqref{eqn_Laplace} that Eq.\ \eqref{eqn_Greens2id} reduces to 
\begin{align}
4\pi \phi({\bf r},t) = \int_{\Omega} \left[ \grad_{\tilde{\bf r}}  \cdot (\phi \grad_{\tilde{\bf r}} G) - \grad_{\tilde{\bf r}} \cdot (G \grad_{\tilde{\bf r}} \phi) \right] d^3 {\bf \tilde{\bf r}}.
\label{eqn_redGreens}
\end{align}
The integral on the right-hand side can be transformed into a surface integral. For closed vortex filaments, the velocity field decays sufficiently fast that if we recede the outer spherical surface, $S^o$, by taking $R \rightarrow \infty$, the outer surface integral vanishes.
We can then write
\begin{align}
\int_{\Omega} \grad_{\tilde{\bf r}} \cdot (\phi \grad_{\tilde{\bf r}} G) d^3 \tilde{\bf r} = \left( \int_{S^+} + \int_{S^-} + \int_{S_T} \right) \phi \grad_{\tilde{\bf r}} G \cdot d\tilde{\bf S}. \label{eqn_surfint}
\end{align}
Across the surface $S(t)$, $G$ is well behaved. Making use of the jump condition given by Eq.\ \eqref{eqn_jumpcondphi}, we obtain
\begin{align}
\left( \int_{S^+} + \int_{S^-} \right) \phi \grad_{\tilde{\bf r}} G \cdot d \tilde{\bf S} = \Gamma \int_{S(t)} \grad_{\tilde{\bf r}} G \cdot d\tilde{\bf S},
\end{align}
where we have made use of the fact that $d\tilde{\bf S}$ is the outward pointing surface element.

It remains to consider the surface integral $\int_{S_T} \phi \grad_{\tilde{\bf r}} G \cdot d\tilde{\bf S}$ in the limit ${\bf r} \rightarrow {\bf s}_i$. 
For the surface integral $S_T$, we note that $d\tilde{\bf S} = -\hat{{\bf e}}_r \epsilon d\tilde{\sigma} d\tilde{\theta}$ where $\hat{{\bf e}}_r$ is a unit vector directed radially away from the filament and into the region $\Omega$, $\epsilon$ is the radius of the tube $S_T$, and $\tilde{\theta}$ is the local angle measured in the plane of the unit vectors ${\bf n}$ and ${\bf b}$.
To proceed, we decompose the integral into two parts
\begin{align}
\int_{S_T} \phi \grad_{\tilde{\bf r}} G \cdot d\tilde{\bf S} = \int_{S_T \backslash S_T'} \phi \grad_{\tilde{\bf r}} G \cdot d\tilde{\bf S} +
\int_{S_T'} \phi \grad_{\tilde{\bf r}} G \cdot d\tilde{\bf S}
\end{align}
where $S_T'$ is a segment of the surface $S_T$ that excludes a small section parameterised by
$|\tilde{\sigma}-\tilde{\sigma}| \le H$ and that contains the point given by ${\bf r}(\sigma) = {\bf s}_i$. Therefore, the first integral on the right hand side contains the singularity as $\epsilon \rightarrow 0$. However, since the integrand under the second integral remains finite, the integral vanishes as the tube is shrunk down to a filament. To evaluate the contribution from the first integral on the right hand side, we proceed to analyse this term by expressing the integrand in terms of local cylindrical coordinates given by $(r,\tilde{\theta},\tilde{\sigma})$ where $r$ and $\tilde{\theta}$ lie in the plane spanned by the principal unit normal and unit binormal vectors. For small $\epsilon$ and taking a sufficiently small segment of the tube, curvature effects of the tube axis can be neglected and the only important nonzero contribution corresponds to
\begin{align}
G = \frac{1}{\left[ r^2 + (\sigma-\tilde{\sigma})^2 \right]^{1/2}}\, , \;\;\;\; 
\frac{\partial G}{\partial r} = \frac{-r}{\left[ r^2 + (\sigma-\tilde{\sigma})^2 \right]^{3/2}} \, .
\end{align}
It follows that since the function $\phi$ locally corresponds to that of a line vortex, we have
\begin{align}
\int_{S_T \backslash  S_T'} \phi \grad_{\tilde{\bf r}} G \cdot d\tilde{\bf S} = \int_{-H}^{H} \int_0^{2\pi} \frac{-\epsilon^2 \Gamma \tilde{\theta} d\tilde{\theta} d\tilde{\sigma} }{(2\pi)\left[ \epsilon^2 + (\sigma-\tilde{\sigma})^2 \right]^{3/2}} \, . 
\end{align}
Focusing on the integral with respect to $\tilde{\sigma}$, we first rescale our coordintaes by setting $\sigma-\tilde{\sigma} = \epsilon \xi$.
In the limit as $\epsilon \rightarrow 0$, the integral over $\tilde{\sigma}$ then transforms to
\begin{align}
\int_{-\infty}^{\infty} \frac{d\xi}{[1+(\xi)^2]^{3/2}} = 2.
\end{align}
We, therefore, recover a finite constant value to the integral for the small segment of the tube enclosing the singular integrand. However, since $\phi$ is determined up to an arbitrary overall constant, we can neglect this term.

Now considering the remaining term on the right-hand side of Eq.\ \eqref{eqn_redGreens}, we have
\begin{align}
\int_{\Omega} \grad_{\tilde{\bf r}} \cdot (G \grad_{\tilde{\bf r}} \phi) d^3 \tilde{\bf r} = \left( \int_{S^+} + \int_{S^-} + \int_{S_T} \right) G \grad_{\tilde{\bf r}} \phi \cdot d\tilde{\bf S}. \label{eqn_surfint}
\end{align}
Since both $G$ and $\grad \phi$ are assumed to be single-valued across $S(t)$ (the velocity around the filament is continuous), the integrals over $S^+$ and $S^-$ cancel one another.
It can be shown that the integral over $S_T$ also vanishes if we approximate the velocity locally as being that of a straight vortex filament
since the flow on the surface of the tube $S_T$ is in the circumferential direction whereas the normal to the surface is in the radial direction in the limit as $\epsilon \rightarrow 0$. 
The expression for the velocity potential of a single closed vortex then reduces to 
\begin{align}
\phi({\bf r},t) = \frac{\Gamma}{4\pi} \int_{S(t)} \grad_{\tilde{\bf r}} \left( \frac{1}{|{\bf r} - \tilde{\bf r}|} \right) \cdot d\tilde{\bf S}, \;\;\;\;\; {\bf r} \notin S(t),
\label{eqn_velpot}
\end{align}
remembering that the definition of the boundaries of the surface $S(t)$ are to be understood in terms of the limiting process obtained from $\epsilon \rightarrow 0$. Equation \eqref{eqn_velpot} shows that the distinguished Seifert surface denoted by $S(t)$ determines the velocity potential over the entire flow. 

To calculate the helicity in terms of the velocity potential, we must recover the velocity potential along the filament.
However, the gradient of the velocity potential which provides an expression for the velocity field assumes the familiar form of the Biot-Savart law upon application of Stokes' theorem to convert the surface integral over $S(t)$ to a contour integral. Extending Eq.\ \eqref{eqn_velpot} to $N$ closed vortex filaments, we obtain
\begin{align}
{\bf v}_{_I}({\bf r},t) = \grad \phi &= \frac{\Gamma}{4\pi} \int_{S(t)} \grad_{{\bf r}} \grad_{\tilde{\bf r}} \left( \frac{1}{|{\bf r}-\tilde{{\bf r}}|} \right) \cdot d\tilde{\bf S} \nonumber \\ 
  &= \sum_{i=1}^N \lim_{\epsilon_i \rightarrow 0} \frac{\Gamma_i}{4\pi}  \oint_{{\mathcal C}_i^*} \grad_{{\bf r}} \left( \frac{1}{|{\bf r} - {\bf s}_i^*(\sigma_i^*,t)|} \right) \times \frac{\partial {\bf s}_i^*}{\partial \sigma_i^*} d\sigma_i^*. \label{eqn_velpotBS}
\end{align}
We note that since Stokes' theorem can only be applied for a smooth function, we take the limit after converting the surface integral to a contour integral. The curve ${\mathcal C}_i^*$ is, therefore, a curve that is arbitrarily close but distinct from ${\mathcal C}_i$ and approaches the latter as $\epsilon_i \rightarrow 0$.

Combining Eqs.\ \eqref{eqn_vort}, \eqref{eqn_Helicity} and \eqref{eqn_velpotBS} we arrive at the final form for the expression of the helicity which is now given by
\begin{align}
{\mathcal H} = \sum_{i=1}^{N} \sum_{j=1}^{N} \lim_{\epsilon_i \rightarrow 0} \frac{\Gamma_i \Gamma_j}{(4 \pi)} \oint_{{\mathcal C}_i} \oint_{{\mathcal C}_j^*} \frac{ ({\bf s}_i - {\bf s}_j^*) \cdot (d{\bf s}_i \times d{\bf s}_j^*)}{|{\bf s}_i-{\bf s}_j^*|^3}. \label{eqn_HelicityCCstar}
\end{align}
We note that the curve ${\mathcal C}_i^*$ is determined by the position of the seam between the slab enclosing the surface $S(t)$ and the surface of the tube $S_{T,i}$ enclosing the filament. Therefore, ${\mathcal C}_i^*$ signifies the local orientation of the distinguished Seifert surface relative to the filament ${\mathcal C}_i$, and it is the twisting of this surface along the filament that determines the twist contribution to the helicity. This result demonstrates that the Seifert framing is the unique correct choice for defining the spanwise vector ${\bf N}_i$ for a vortex. 

We remark that associating the twist contribution to the local phase structure in the vicinity of the filament is also consistent with similar results discussed previously in the context of wave dislocations in optical phase fields \cite{Dennis2004,Dennis2009}.
However, it is essential to recognise that in contrast to a vortex tube where the twist contribution arises from additional degrees of freedom associated with the distribution of the vorticity within a vortex core, no such freedom exists within a filament. Therefore, the instantaneous configuration of vortices determines the twist contribution. In contrast, the twist of a vortex tube is not completely prescribed by the configuration of the centreline of the vortex tube. 

The integrals appearing in Eq.~\eqref{eqn_HelicityCCstar} are of the form considered by Calugareanu \cite{Calugareanu1959,Calugareanu1961}. 
To evaluate the different contributions to helicity, we first consider the contour integral along the curve ${\mathcal C}_i$. To isolate the singular part of the integrand arising from ${\bf s}_i \rightarrow {\bf s}_i^*$ as $\epsilon_i \rightarrow 0$, we consider a small arc, denoted by $\mathcal{I}$ that contains the point ${\bf s}_i(\sigma_i)$ with limits denoted by ${\bf s}(\sigma_i-H)$ and ${\bf s}(\sigma_i+H)$. For brevity, we will drop the subscript, $i$, in what follows. The integral over ${\mathcal C}$ is then given by
\begin{align}
\left( \int_\mathcal{I} + \int_{{\mathcal C} \backslash  \mathcal{I}} \right) \frac{ ({\bf s} - {\bf s}^*) \cdot (d{\bf s} \times d{\bf s}^*)}{|{\bf s}-{\bf s}^*|^3}.
\label{eqn_Helicity_reg}
\end{align}
We now focus on the first integral and consider what happens as we let $\epsilon \rightarrow 0$ with $H$ fixed. We can parameterise the curve ${\mathcal C}^*$ in terms of the arclength $\tilde{\sigma}$ along the curve ${\mathcal C}$ by setting 
$\mathbf{s}^* = \tilde{\mathbf{s}} + \epsilon(\tilde{{\bf n}} \cos \Theta(\tilde{\sigma})+\tilde{{\bf b}} \sin \Theta(\tilde{\sigma}))$
where $\Theta(\tilde{\sigma})$ determines the direction of the spanwise vector in the plane of the unit principal normal and unit binormal vectors. The angle $\Theta$ is measured relative to the normal and is increasing in the counterclockwise direction for a right-handed coordinate system. We note that an arclength parameterisation $\sigma$ for the curve ${\mathcal C}$ is not necessarily so for the curve ${\mathcal C}^*$. Therefore, in evaluating terms contributing to the second integral of Eq.\ \eqref{eqn_Helicity_reg}, we find
\begin{align}
{\bf t}^* \frac{\partial \sigma^*}{\partial \tilde{\sigma}} = \tilde{{\bf t}} (1-\epsilon \kappa(\tilde{\sigma}) \cos \Theta) + \tilde{{\bf n}} \left[ - \epsilon (\tau(\tilde{\sigma}) + \Theta') \sin \Theta \right] + \tilde{{\bf b}} \left[ \epsilon (\tau(\tilde{\sigma}) + \Theta') \cos \Theta \right],
\end{align}
where ${\bf t}^* \equiv \frac{\partial {\bf s}^*}{\partial \sigma^*}$. Moreover, by introducing the small parameter $h=\sigma-\tilde{\sigma}$ and expanding using Taylor's series, we have
\begin{align}
{\bf s} - \tilde{{\bf s}} &= \tilde{{\bf t}} (h+h^3 o(1)) + \tilde{{\bf n}} \left( \frac{h^2}{2}\kappa(\tilde{\sigma})+h^3 o(1) \right) + \tilde{{\bf b}} h^3 o(1) , \nonumber \\
{\bf t} &= \tilde{{\bf t}} (1+h^2 o(1)) + \tilde{{\bf n}} \left( h\kappa(\tilde{\sigma}) +h^2 o(1) \right) + \tilde{{\bf b}} h^2 o(1) . \nonumber
\end{align}
Using these relations, we can express the numerator of the integrand as
\begin{align}
({\bf s} - {\bf s}^*) \cdot \left(d{\bf s} \times d{\bf s}^* \right) =& \left[ \epsilon^2 (\tau+\Theta')  +\epsilon h \kappa \sin \Theta  \right. \nonumber \\
& \left. - \epsilon^2 h \kappa^2 \sin \Theta \cos \Theta + h^4 o(1) + \epsilon h^2 O(1) \right] d\sigma d\tilde{\sigma}^* ,
\label{eqn_sdiff_tdiff}
\end{align}
whereas the denominator is given by
\begin{align}
|{\bf s} - {\bf s}^*|^3 = (h^2 + \epsilon^2 - \epsilon h^2 \kappa \cos \Theta + h^3 o(1))^{3/2}. \label{eqn_sdiff_abs}
\end{align}
We can now see from these expressions that $\lim_{h\rightarrow 0} \lim_{\epsilon \rightarrow 0} (\cdot) =0$ whereas $\lim_{\epsilon \rightarrow 0} \lim_{h \rightarrow 0} (\cdot) = (\tau + \Theta')$. This is an example of a non-uniformly convergent expression and is the essential reason why a naive derivation of the expression for the helicity as given in Eq.\ \eqref{eqn_HelicityBS} fails to capture the twist contribution. According to Eq.\ \eqref{eqn_HelicityCCstar}, the limit with respect to $\epsilon$ should not be taken first. To proceed in a more systematic way, we will consider a distinguished limit obtained by setting $h = \epsilon s$ after which the integral reduces to the form
\begin{align}
I &= \lim_{\epsilon \rightarrow 0} \int_{-H/\epsilon}^{H/\epsilon} \frac{f(s) ds}{(1+s^2)^{3/2}} + O(\epsilon) = \frac{1}{2\pi} (\tau + \Theta'), \nonumber \\
f(s) &= (\tau + \Theta')+As +\epsilon Bs^2+\epsilon Cs + \epsilon^2 D s^2 + \epsilon^2 E s^4 . 
\label{eqn_I_integral} 
\end{align}
$A$, $B$, $C$, $D$, and $E$ are coefficients related to $\Theta$ and $\kappa$ evaluated at the point $\sigma=\tilde{\sigma}$. For fixed $H$, the limits tend to $\pm \infty$ whereas terms regular in $\epsilon$ vanish. Since the above integral was analysed in detail by Calugareanu\cite{Calugareanu1959,Calugareanu1961} we have not rederived the above result here. However, to expound on the arguments in those works, we have included details in the Appendix.

Hence, we have recovered the expression for the helicity for a vortex filament that contains a nontrivial contribution arising from the torsion. We end this section by noting that a consequence of working in the Seifert framing is that the continuum definition of the helicity that we have used must be zero.
We begin by noting that
\begin{align}
\lim_{\epsilon_i \rightarrow 0} \frac{\partial}{\partial \sigma_i^*}({\bs}_i^* \cdot  \left. \grad \phi \right|_{\br=\bs_i^*}) = \lim_{\epsilon_i \rightarrow 0} \left( \frac{\partial \bs_i^*}{\partial \sigma_i^*} \cdot  \left. \grad \phi \right|_{\br=\bs_i^*} + {\bs}_i^* \cdot  \frac{\partial}{\partial \sigma_i^*}(\left. \grad \phi \right|_{\br=\bs_i^*}) \right) \, .
\label{eqn_sdotgradphi}
\end{align}
By construction, the phase along the curve $\mathcal{C}_i^*$ is a constant $\phi(\bs_i^*(\sigma_i^*))=\phi_0$. 
This implies that the second term in Eq.\ \eqref{eqn_sdotgradphi} vanishes. Since ${\bf v}_{_I}(\br) = \grad \phi(\br)$, it follows that if we use Eq.\ \eqref{eqn_BS} and integrate Eq.\ \eqref{eqn_sdotgradphi} along the contour $\mathcal{C}_i^*$ we obtain
\begin{align}
\sum_{j=1}^N \lim_{\epsilon_i \rightarrow 0}  \oint_{\mathcal{C}_i^*}  \oint_{\mathcal{C}_j} \frac{{\partial \bf s}_i^*}{\partial \sigma_i^*}  \cdot \frac{({\bs}_j -\bs_i^*(\sigma_i^*)) \times d{\bs}_j}{|{\bs}_j-\bs_i^*(\sigma_i^*)|^3} d\sigma_i^* = 0. 
\end{align}
The above expression is zero because we are integrating an exact differential on the left hand side of Eq.\ \eqref{eqn_sdotgradphi}. It follows that the helicity $\mathcal{H} = 0$.

\begin{figure}[t]
\makeatletter
  \def\@captype{figure}
\makeatother
\centering
 \begin{minipage}[b]{1.\textwidth}  
 \begin{minipage}[b]{0.5\textwidth} 
  \centering
  \subfigure[Top view]{
\includegraphics[scale=0.34]{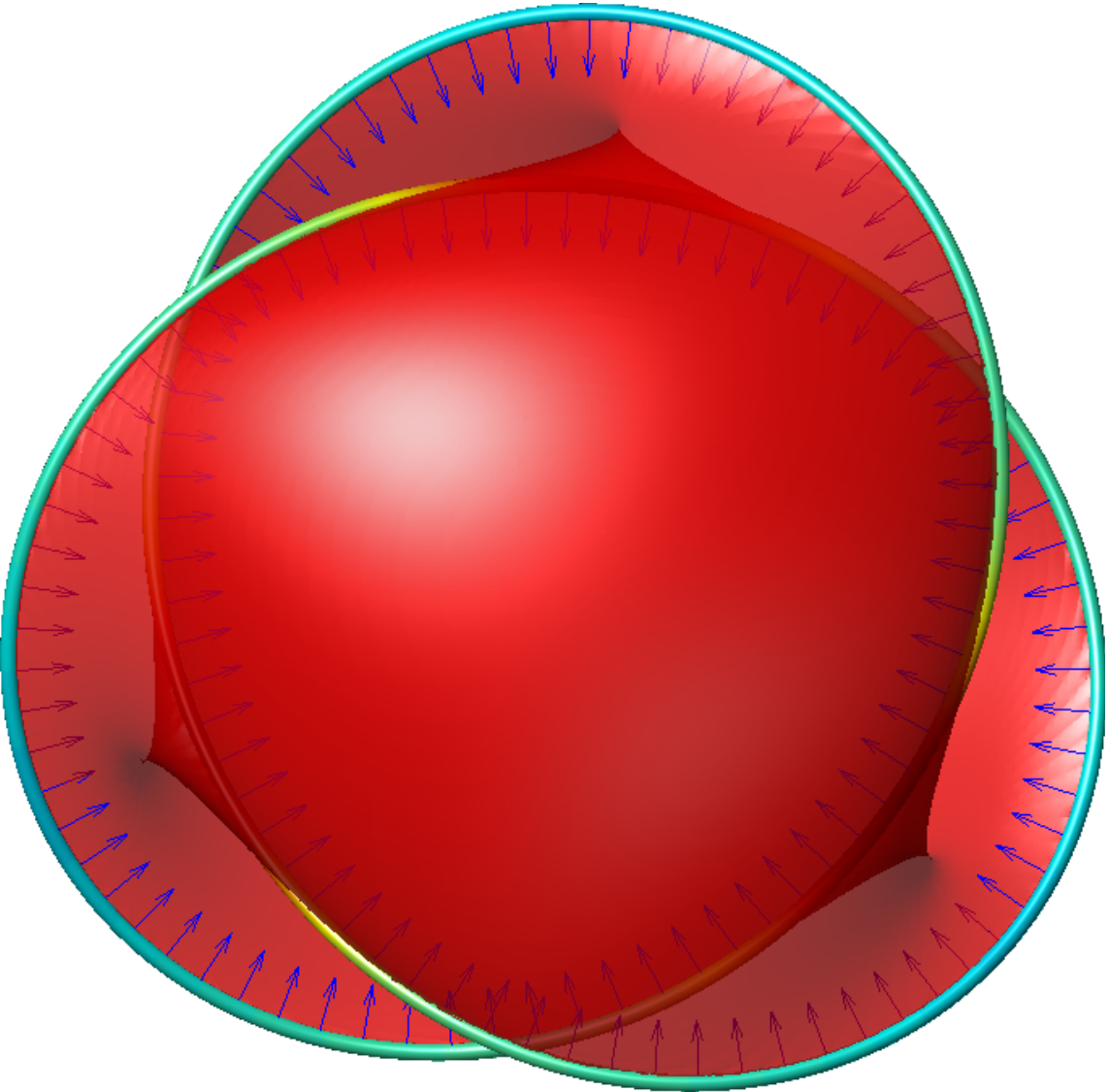}}
 \end{minipage}%
 \begin{minipage}[b]{0.5\textwidth} 
  \centering
   \subfigure[Side view]{
\includegraphics[scale=0.3]{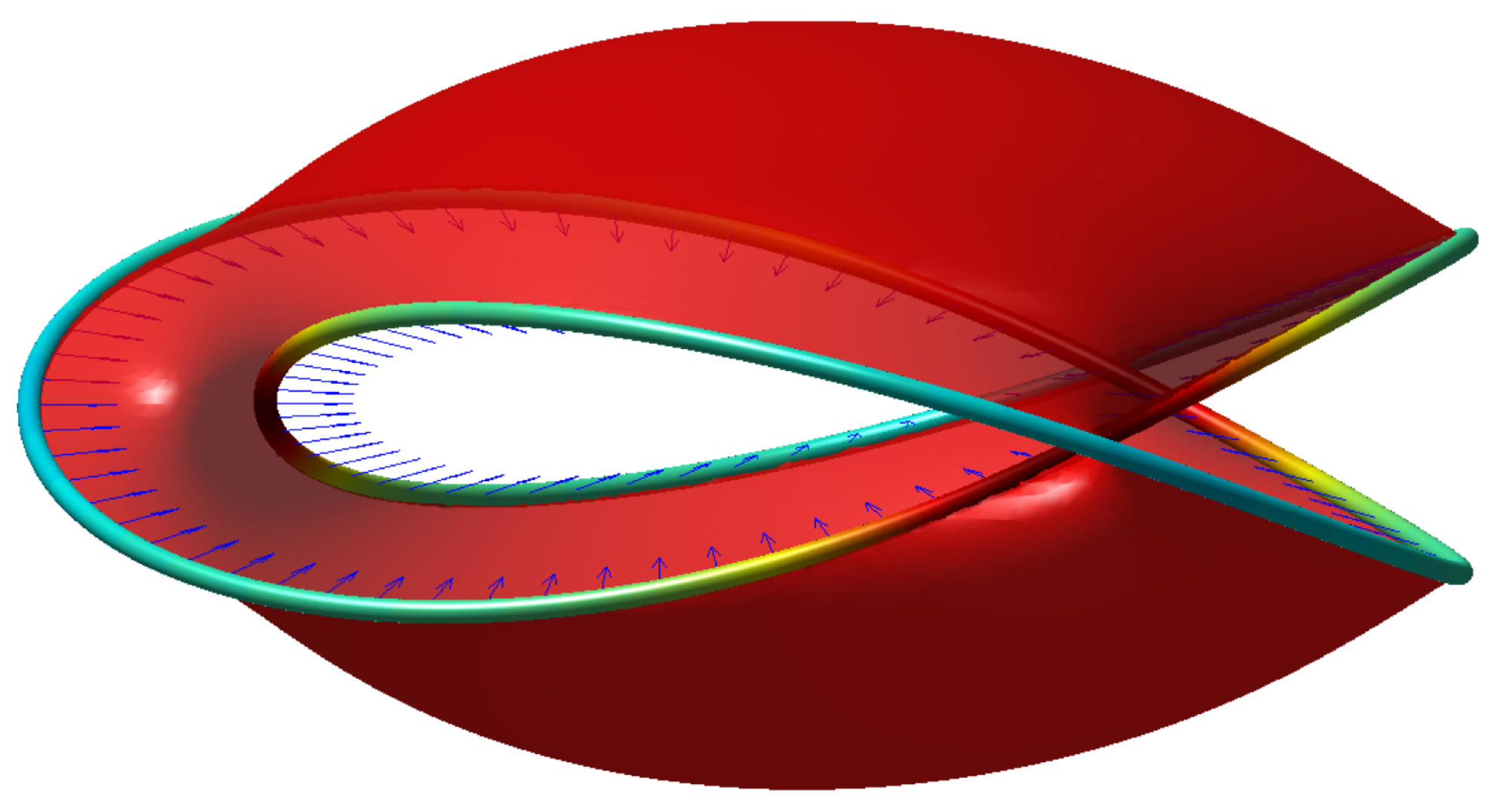}}
  \end{minipage}   
  \end{minipage}     
  \vspace{0.1cm}
  \caption{Illustration of a distinguished Seifert surface for a trefoil knot. The surface is an orientable smooth surface with its boundary delineated by the knot. Such a surface can always be defined for any smooth (possibly knotted) curve. The color coding on the tube corresponds to the value of the local torsion whereas the arrows along the filament are pointing in the direction of the local normal vector; (a) Top view and (b) side view. \label{Fig_Seifert}}
\end{figure}

To clarify how a smooth Seifert surface spanning a knotted vortex can be constructed, we have computed the velocity potential corresponding to a trefoil vortex described by the form given in \cite{Ricca1993}. The coordinates of the trefoil are given by
\begin{align}
x &= r \cos (\alpha) , \;\;\;\; y= r \sin (\alpha), \;\;\;\; z = \left[1 - \left( {p}/{q} \right)^2 \right]^{1/2} \cos (q \theta) \, , \;\;\;\;\; \nonumber \\
r &= r_0 + a \sin (q \theta), \;\;\;\;\;
\alpha = p \theta + \frac{p}{q} \frac{a}{r_0} \cos(q \theta) \, , \label{eq_knot}
\end{align}
where $\theta \in [0,2\pi)$. The velocity potential is constructed by using the approaches described in \cite{Salman2013,Scheeler2014}. The specific parameters chosen for the trefoil shown in Fig.\ (2) are $p=2$, $q=3$, $r_0 = 28$, and $a = 5$. The normal vector for the trefoil is illustrated in Fig.\ (\ref{Fig_Seifert}) together with the corresponding Seifert surface. As is evident, despite the non-trivial topology, the surface smoothly subtends the vortex filament.

\section{Quasiclassical Limit}

A trivial result for helicity raises the question, `what is its physical significance for superfluids?'. An immediate consequence of the above result is that the sum of the total self-linking ($SL$) number of all curves (sum of writhe and twist contributions) is equal in magnitude, but opposite in sign, to the sum of the linking number ($Lk$). This gives
\begin{align}
Lk +SL = 0,
\end{align}
implying that there is a duality between linking and self-linking. Now, it is well known that if superfluids are forced at large scales then large amounts of polarized vorticity are injected into the superflow \cite{Walmsley2001}. These polarized bundles have been invoked in explaining aspects of quasiclassical turbulence in superfluids that experimentally have been found to exhibit a Kolmogorov like regime at low wavenumbers.

An important consequence of the presence of polarized vortex bundles is that if a coarse-grained description of the vortex lines is adopted, we recover a smooth non-trivial vorticity. In particular, if we consider the scenario of many vortex lines concentrated in bundles that are otherwise separated by irrotational flow, then it is conceivable that a control volume of fluid, $\mathcal{D}$, with volume $V_{\mathcal{D}}$ can be threaded by many of these vortex lines. Rather than track the positions of individual lines, we can then define a coarse-grained vorticity vector as
\begin{eqnarray}
\bomega_s = \frac{1}{V_{\mathcal{D}}} \sum_{i=1}^N \Gamma_i \int_{\mathcal{D}} \oint d\sigma_i \frac{\partial {\bf s}_i}{\partial \sigma_i} \delta^{(3)} \left( {\bf r} - {\bf s}_i(\sigma_i,t) \right) d^3{\bf r}.
\end{eqnarray}
In this scenario, this coarse-grained description would, to a good approximation, correspond to vortex tubes with a smooth vorticity in an otherwise irrotational flow. This is in contrast to a tangle of vortex lines with no particular preferred orientation. Since the coarse-graining procedure produces a rotational flow within the vortex tube, a velocity potential can not be defined within the core of the vortex. Consequently, we lose all information about the self-linking of the vortices with the velocity potential. On the other hand, the information about the linking numbers between the curves is contained in a coarse-grained description in the smooth vorticity field within the tube. Therefore, in the quasiclassical regime, that coincides with vortex bundles, the coarse-graining procedure gives rise to a non-zero continuum helicity at a macroscopic level. It follows, that only the linking numbers are relevant in order to relate the topology of the vortex lines on the microscopic level to the macroscopic helicity. This is also consistent with the definition of helicity for discrete vortex filaments given by Moffatt (1969). Our results, therefore, clarify the relation between the definitions of helicity for continuum vorticity fields and discrete vortex filaments. We recall that in contrast to writhe or torsion, the linking numbers always correspond to integral values. Hence, only these well-behaved components of helicity on the microscopic level determine the macroscopic description. 

We remark that a particular example of the coarse-graining procedure discussed above leads to the Hall-Vinen-Bekarevich-Khalatnikov equations, in which the vorticity of the superfluid component is described by a continuous field $\bomega_s = \text{curl} {\bf v}_s$. The Hall-Vinen-Bekarevich-Khalatnikov (HVBK) equations \cite{Hall1956,Bekarevich1961,Barenghi2001} provide a generalization of Landau's two-fluid model of superfluid Helium by including the effect of superfluid vortices that couples the superfluid and normal fluid components. It provides a macroscropic description of the superflow in the sense that a control volume of the fluid is assumed to be large enough that it is threaded by many vortices. A key working assumption of the model is that the individual superfluid vortices are locally aligned with one another thus permitting a coarse-grained vorticity field $\bomega_s$ to be defined. The model, is therefore, appropriate in describing the flow on length scales much larger than the intervortex separation. If compressibility is neglected, the HVBK equations can be expressed as 
\begin{align}
\frac{\partial \bv_n}{\partial t} + (\bv_n \cdot \bnabla) \bv_n &= -\frac{1}{\rho}\bnabla P - \frac{\rho_s}{\rho_n} S \bnabla T + \nu_n \nabla^2 \bv_n + \frac{\rho_s}{\rho} {\bf F}, \nonumber \\
\frac{\partial \bv_s}{\partial t} + (\bv_s \cdot \bnabla) \bv_s &= -\frac{1}{\rho}\bnabla P + S\bnabla T + {\bf T} - \frac{\rho_n}{\rho} {\bf F}, 
\end{align}
where the mututal friction ${\bf F}$, and tension force ${\bf T}$, are given by
\begin{align}
{\bf F} &= \frac{B}{2} \hat{\bomega}_s \times \big[ \bomega_s \times \bs_0 \big] + \frac{\tilde{B}}{2} \bomega_s \times \bs_0 , \nonumber \\
{\bf T} &= -\nu_s \bomega_s \times (\bnabla \times \hat{\bomega}_s), \;\;\;
\nu_s = (\Gamma/4\pi) \ln(b_0/a_0),
\end{align}
and  
\begin{align}
\bomega_s = \grad \times {\bv}_s, \;\;\; \hat{\bomega}_s = \bomega_s/|\hat{\bomega}_s|, \;\;\; \bs_0 = (\bv_n-\bv_s-\nu_s \bnabla \times \hat{\bomega}_s).\nonumber 
\end{align}
Here, $\rho_s$, $\rho_n$, $\bv_s$, $\bv_n$, $T$, $S$, $\nu_n$, $\nu_s$, $b_0$, $a_0$, $B$, and $\tilde{B}$  are the superfluid density, normal fluid density, superfluid velocity, normal fluid velocity, temperature, entropy, kinematic viscosity, the vortex tension parameter, the intervortex spacing, the cut-off length scale related to the vortex core size, and the two mutual friction coefficients, respectively.

The above form of the equations clearly reveal the close analogy between the equation for the superfluid velocity and the classical Euler equations. Although the mutual friction ${\bf F}$ and the tension force ${\bf T}$ act as source terms, in the limit as the temperature tends to zero, we have $\rho_n \rightarrow 0$. Moreover since $\nu_s \sim \kappa$ which is set by Planck's constant, we can ignore this term, in which case an equation for a pure superflow is recovered. Under such conditions the helicity conservation theorems of classical fluids are expected to hold and, hence, when vortex bundles are considered, the microscopic form of the helicity for a quantum system must coincide with the quasi-classical form. In this case a non-zero quasiclassical helicity given by $\mathcal{H}_{cl} = \int {\bf v}_s \cdot \bomega_s$ is then recovered.

The dynamics of vortex bundles including their mutual interactions during reconnections was considered in \cite{Alamri2008}. It was found that the coherence of the bundles persists following reconnections. Recently, conservation of the helicity  of vortex bundles in the quasiclassical regime was studied in \cite{Hanninen2016} with the vortex filament bundle. This confirmed that, to a good approximation, quasiclassical helicity is conserved in the absence of reconnections provided the bundle retains its coherence. These results are consistent with the identification of quasiclassical helicity with the linking number of vortex lines within a bundle. Moreover, vortex bundles with non-trivial topology can also reconnect while preserving their coherence. To demonstrate this, we have used the Gross-Pitaevskii model of a superfluid to study how an initial configuration consisting of two linked vortex bundles in the form of a Hopf-link relaxes with time. The initial condition for the Hopf-links was constructed using two bundles with each bundle containing 7 vortices. The coordinates of the central vortices in each bundle were given by
\begin{align}
x_1 &= r \cos (\theta)-20.25, \;\;\;\;\;  y_1 = r \sin (\theta) +19.5, \;\;\;\;\;
z_1 = 0 \, , \nonumber \\
x_2 &= r \cos (\theta)+20.25, \;\;\;\;\;  y_2 = 0.,  \;\;\;\;\;
z_2 = r \sin (\theta) \, ,
\end{align}
where $r=40.5$. The surronding six vortices were then distributed uniformly around each central vortex by offsetting them at a radial distance of 4 from each central vortex.

We integrated the non-dimensional form of the Gross-Pitaevskii equation given by
\begin{align}
i\psi_t = -\nabla^2 \psi + |\psi|^2\psi,
\end{align}
where $\psi$ is the macroscopic complex wavefunction of the superfluid. We discretised our 3D periodic domain on a grid consisting of $512^3$ points along the three coordinate directions with a grid spacing of $\Delta x = \Delta y = \Delta z = 0.5$. The equations were then integrated in time using a fourth order split-step scheme in which the linear term is integrated forward in time in Fourier space whereas the nonlinear term is integrated in physical space.
The equations were integrated using a timestep of $\Delta t =0.1$.
\begin{figure}[t]
\makeatletter
  \def\@captype{figure}
\makeatother
\centering
 \begin{minipage}[b]{1.\textwidth}  
 \begin{minipage}[b]{0.5\textwidth} 
  \centering
  \subfigure[Initial condition]{
\includegraphics[scale=0.41]{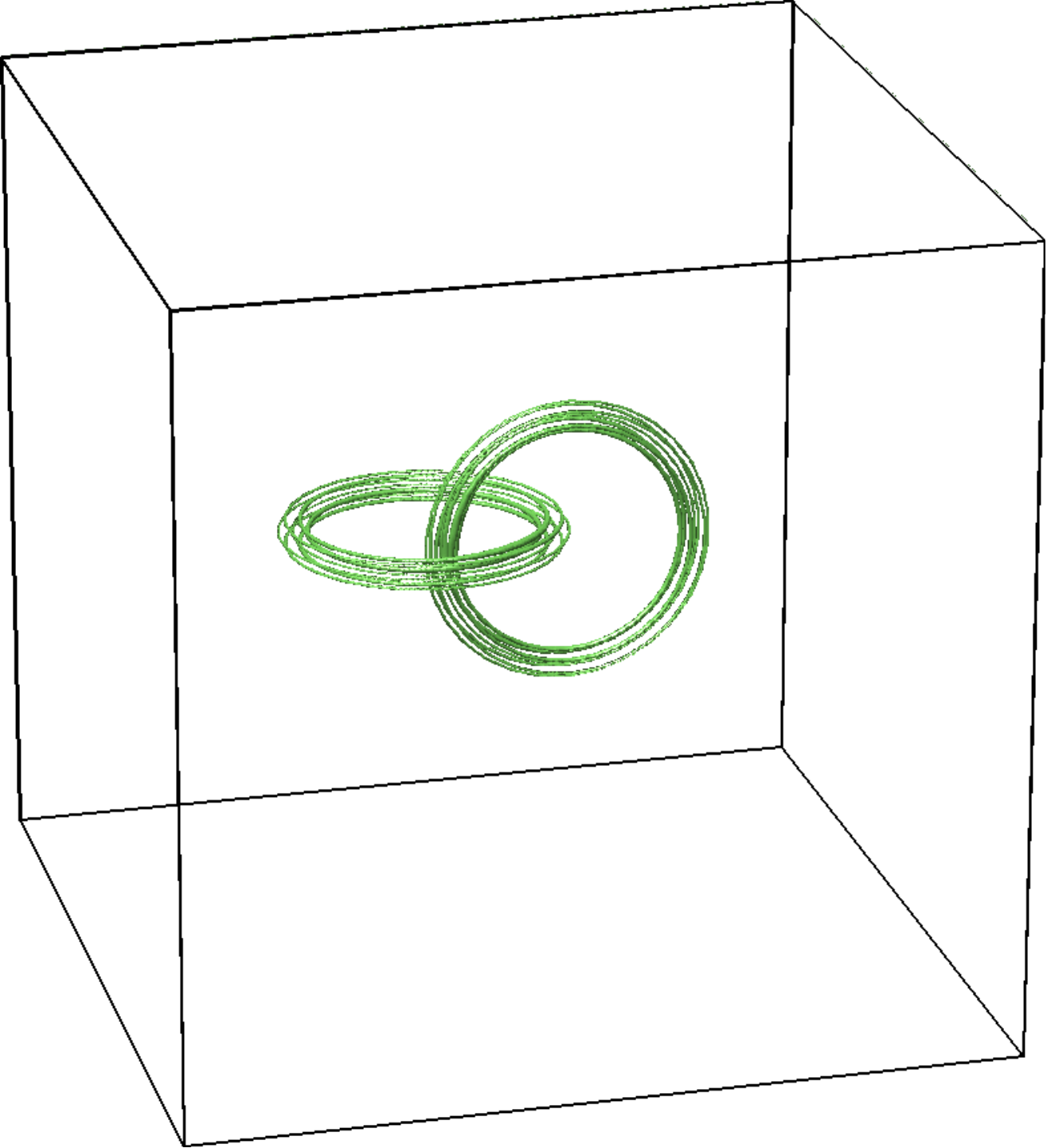}}
 \end{minipage}%
 \begin{minipage}[b]{0.5\textwidth} 
  \centering
   \subfigure[Intermediate configuration]{
\includegraphics[scale=0.41]{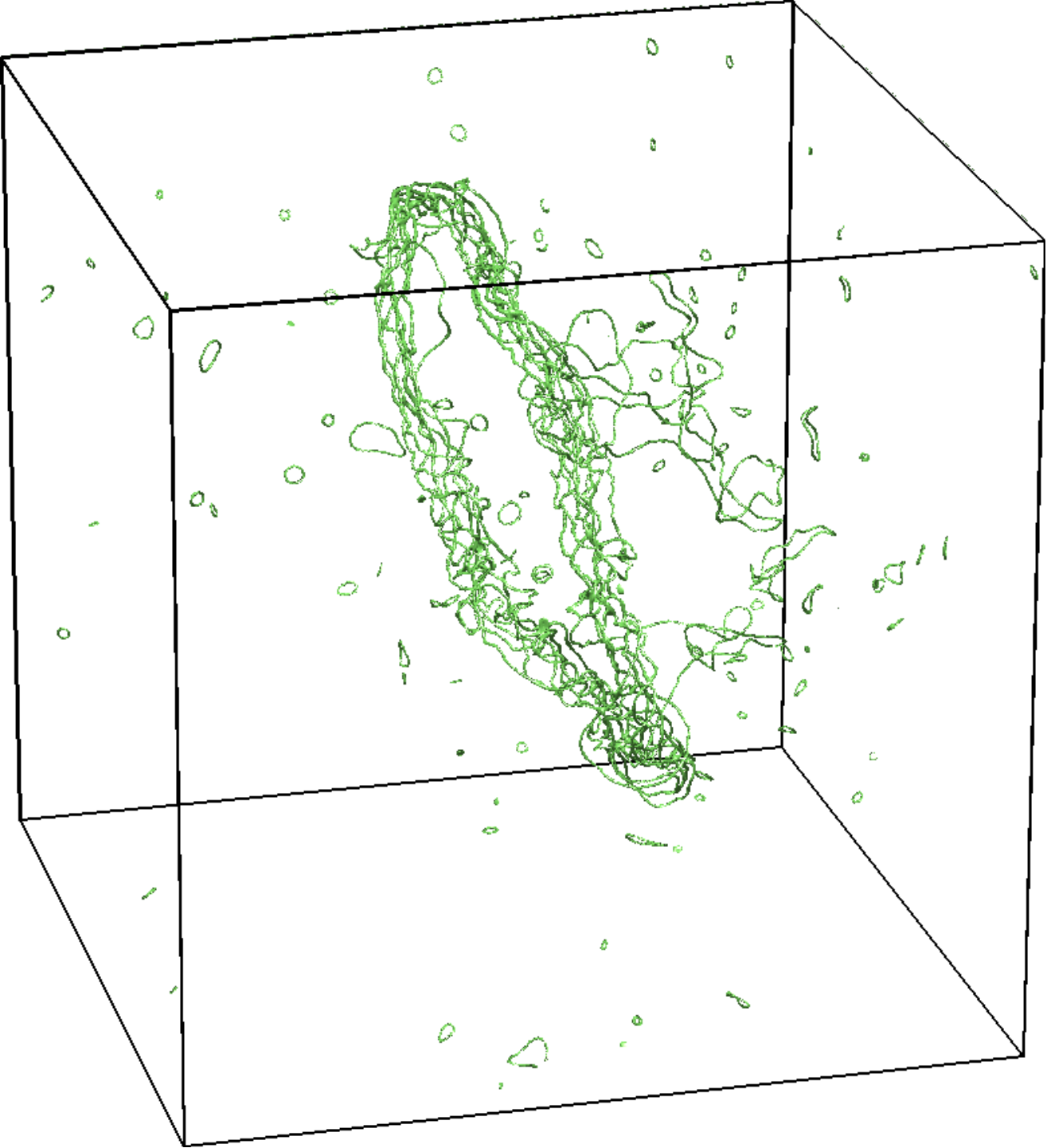}}
  \end{minipage}   
  \end{minipage}   
  \vspace{0.1cm}
  \caption{Configuration of vortex bundles initialised in the form of a hopf link: (a) Initial configuration; (b) Bundle in the form of vortex ring emerging at intermediate time following reconnection. The reconnections produce Kelvin waves on individual vortices as well as the emission of many small rings. \label{fig_bundles}}
\end{figure}
The initial condition is shown in Fig.\ \ref{fig_bundles}a. The configuration of the vortices following the reconnections of the bundles is presented in Fig.\ \ref{fig_bundles}b. As can be seen from the figures, the two bundles reconnect and emerge as one large coherent bundle of vortex rings. At the same time, Kelvin waves are excited on the vortex filaments making up the bundle. Although these Kelvin waves can contribute writhe to the helicity of individual vortices, this writhe does not contribute to the quasiclassical form of helicity which depends only on the linking between the vortices that changes following the reconnections.

\section{Conclusions}

Starting from first principles, we have shown how the Seifert framing emerges from the continuum definition of helicity to provide a unique twist contribution to a superfluid vortex filament. Our results remove an element of arbitraryness that is often associated with the choice of the spanwise vector for superfluid vortices. They show that a Seifert surface associated with a surface of constant velocity potential provides an important contribution to the topological properties of knotted filaments. An immediate consequence of using the Seifert framing is that the continuum definition of helicity is zero and essentially provides a relation between the self-linking and the linking numbers. This duality between the two components of helicity motivates a definition of helicity based on linking number alone which can explain the emergence of a non-trivial quasiclassical helicity in terms of coherent vortex bundles. It, therefore, allows us to reconcile our results for superfluid vortices to properties of helicity in classical fluids.

We end by remarking that it may appear odd to find that a surface of discontinuity of the velocity potential has physical significance even though it is commonly accepted that the phase merely provides a convenient mathematical construct to describe a fluid flow. However, we note that the surface of discontinuity results in measureable effects on waves scattering off a vortex which in the context of superfluids results in the Iordanskii force \cite{Iordanskii1966,Stone2000,Sonin2012}. This is the analogue of the Aharonov-Bohm effect for electrons scattered by a magnetic flux tube. Therefore, it should not be too surprising that careful consideration is required for the twisting of the velocity potential around a vortex filament when attempting to evaluate the continuum form of the helicity.\\






\acknowledgments{The author acknowledges support for a Research Fellowship from the Leverhulme Trust under Grant R201540. The author would like to thank Risto H\"{a}nninen and Niklas Hietala for many helpful discussions.}

\appendix
\section{Derivation of Twist Contribution}

In order to recover the twist contribution, we follow the arguments presented in \cite{Calugareanu1961}. Firstly, the numerator and denominator of Eq.\ \eqref{eqn_Helicity_reg} are evaluated with the use of Eqs.\ \eqref{eqn_sdiff_tdiff}-\eqref{eqn_sdiff_abs}. For the numerator of the integrand, we can write
\begin{align}
({\bf s} - {\bf s}^*) \cdot (d{\bf s} \times d{\bf s}^* ) = ({\bf s} - {\bf s}^*) \cdot \left({\bf t} \times {\bf t}^* \right) \frac{\partial \sigma^*}{\partial \tilde{\sigma}} d\sigma d\tilde{\sigma}.
\end{align}
Since the triple scalar vector product is given by ${\bf c} \cdot ({\bf d} \times {\bf e}) = c_3(d_1e_2-d_2e_1)+c_2(d_3e_1-d_1e_3)+c_1(d_2e_3-d_3e_2)$, setting ${\bf c} \cdot ({\bf d} \times {\bf e})= ({\bf s} - {\bf s}^*) \cdot ({\bf t} \times d{\bf s}^*/d\tilde{\sigma})$ gives
\begin{align}
c_1(d_2e_3-d_3e_2) =& (h+h^3o(1)) \big[ \epsilon (\tau + \Theta') \cos \Theta  (k\kappa+h^2o(1)) + \epsilon (\tau + \Theta') \sin \Theta h^2 o(1) \big],
\nonumber \\
c_2(d_3e_1-d_1e_3) =& -\left[ \frac{h^2}{2} \kappa - \epsilon \cos \Theta + h^3 o(1) \right] \big[ \epsilon (\tau + \Theta') \cos \Theta (1+h^2o(1)) 
- (1-\epsilon \kappa \cos \Theta) h^2 o(1) \big], \nonumber \\
c_3(d_1e_2-d_2e_1) =& \big[ h^3 o(1) - \epsilon \sin \Theta \big] \big[ -\epsilon (\tau + \Theta') \sin \Theta (1+h^2o(1)) 
-(1-\epsilon \kappa \cos \Theta)(h\kappa + h^2o(1)) \big],
\end{align}
where ${\bf c} = (c_1,c_2,c_3)$ etc.\ from which it follows that
\begin{align}
({\bf s} - {\bf s}^*) \cdot \left({\bf t} \times {\bf t}^* \right) \frac{\partial \sigma^*}{\partial \tilde{\sigma}}  &= \left[ \epsilon^2 (\tau+\Theta')  +\epsilon h \kappa \sin \Theta  \right. \\
& \left. - \epsilon^2 h \kappa^2 \sin \Theta \cos \Theta + h^4 o(1) + \epsilon h^2 O(1) \right] \equiv g(h) . \nonumber
\end{align}
Similiarly, for the denominator of Eq.\ \eqref{eqn_Helicity_reg}, we obtain
\begin{align}
|\bs-\bs^*|^3 = (h^2 +\epsilon^2 - \epsilon h^2 \kappa \cos \Theta + h^3 o(1))^{3/2}.
\end{align}
The resulting integral can now be expressed as
\begin{align}
I = \lim_{\epsilon \rightarrow 0} \int_{-H}^{H} \frac{g(h) dh}{(\epsilon^2+h^2)^{3/2} \left[1-\epsilon h^2/(\epsilon^2+h^2) \kappa \cos \Theta + h^3/(\epsilon^2+h^2) o(1) \right]^{3/2} } \, , \nonumber
\end{align}
where $\Theta$ and $\kappa$ are evaluated at the point $\sigma=\tilde{\sigma}$.
We are interested in evaluating the integral as $\epsilon \rightarrow 0$. However, if we let $\epsilon \rightarrow 0$ with $h$ fixed, then we recover
\begin{align}
\frac{h^4 o(1)}{h^2+h^3 o(1)} = h^2 o(1) \rightarrow 0 \;\;\; \textrm{as} \;\;\; h \rightarrow 0,
\end{align}
for the integrand. On the other hand, if we take $h \rightarrow 0$ first then the integrand becomes $(\tau+\Theta')$
and we obtain a finite contribution. This is an example of a non-uniformly convergent limit. We note that the first limit would be applicable when $|\bs-\bs^*| \gg \epsilon$ whereas the second limit should apply when $|\bs-\bs^*| \ll \epsilon$. We will, therefore, consider the distinguished limit obtained by setting $h=\epsilon s$. The integral then becomes
\begin{align}
I &= \lim_{\epsilon \rightarrow 0} \int_{-H/\epsilon}^{H/\epsilon} \frac{f(s) ds}{(1+s^2)^{3/2} \left[1-\epsilon s^2/(1+s^2) \kappa \cos \Theta + \epsilon s^3/(1+s^2) O(1) \right]^{3/2} } , \nonumber \\
&=  \lim_{\epsilon \rightarrow 0} \int_{-H/\epsilon}^{H/\epsilon} \frac{f(s) ds}{(1+s^2)^{3/2} } (1+\epsilon \chi(s)), \nonumber \\
f(s) &= (\tau + \Theta')+As +\epsilon Bs^2+\epsilon Cs + \epsilon^2 D s^2 + \epsilon^2 E s^4, \nonumber
\end{align}
where we have made use of the binomial theorem to arrive at the second line. $A$, $B$, $C$, $D$, and $E$ are coefficients related to $\Theta$ and $\kappa$ evaluated at the point $\sigma=\tilde{\sigma}$. For fixed $H$, the limits tend to $\pm \infty$ whereas terms regular in $\epsilon$ vanish. We can now consider each term of the integral separately.

The integrals corresponding to the coefficients $A$ and $C$ both vanish since the integrals are odd functions of $s$. For $T>1$, which applies when $\epsilon \rightarrow 0$, the integrals involving the $B$ and $D$ coefficients are of the form
\begin{align}
\int_{-T}^T \frac{s^2 ds}{(1+s^2)^{3/2}} &= 2\int_0^1 \frac{s^2 ds}{(1+s^2)^{3/2}}
+ 2 \int_1^T \frac{s^2 ds}{(1+s^2)^{3/2}} \nonumber \\
&< 2\int_0^1 \frac{s^2 ds}{(1+s^2)^{3/2}} +
2 \int_1^T \frac{ds}{s} = 2 \left\{ \frac{-1}{\sqrt{2}} + \ln T \right\}.
\end{align}
Since the two terms are multiplied by $\epsilon$ and $\epsilon^2$, which correspond to the prefactors $1/T$ and $1/T^2$, respectively, the terms containing the $B$ and $D$ coefficients also vanish in this limit. For the final term, containing the $E$ coefficient, we have
\begin{align}
\lim_{T \rightarrow \infty} \frac{1}{T^2} \int_{-T}^{T} \frac{s^4 ds}{(1+s^2)^{3/2}} &= \lim_{T \rightarrow \infty} \frac{2}{T^2} \int_0^T \frac{s^4 ds}{(1+s^2)^{3/2}} \\
&= \lim_{T \rightarrow \infty}  \left( \frac{-2T}{\sqrt{1+T^2}}
+ \frac{6\sqrt{1+T^2}}{T} - \frac{6}{T^2} \int_0^T \sqrt{1+s^2} ds \right) = 1. \nonumber
\end{align}
On the other hand, the first integral involving the coefficient $(\tau+\Theta')$ can be integrated explicitly to obtain
\begin{align}
\lim_{T\rightarrow \infty} \int_{-T}^T \frac{ds}{(1+s^2)^{3/2}} = \lim_{T\rightarrow \infty} 2\int_0^T \frac{ds}{(1+s^2)^{3/2}} = 2.
\end{align}
Combining all of these together leads to the final result given in Eq.\ \eqref{eqn_I_integral}.

\end{document}